\documentclass[preprint,12pt]{elsarticle}

\usepackage{amsmath,amssymb, amsopn}
\usepackage{amsthm}

\usepackage{graphicx}
\usepackage{epstopdf}
\usepackage{algorithmic}
\usepackage{comment}
\usepackage[ruled,lined]{algorithm2e}
\usepackage{url}

\usepackage[table]{xcolor}

\usepackage{array}

\newtheorem{theorem}{Theorem}[section]
\newtheorem{lemma}[theorem]{Lemma}

\newtheorem{definition}{Definition}[section]
\newtheorem{remark}{Remark}[section]

\usepackage{cleveref}

\journal{Information Systems}

\begin{document}
\begin{frontmatter}
\title{Estimating Diffusion Degree on Graph Streams} 

\author[cseiitj]{Vinit Ramesh Gore}
\ead{gore.1@iitj.ac.in}

\author[cseiitj]{Suman Kundu \corref{cor1}} 
\ead{suman@iitj.ac.in}

\author[cseiitj]{Anggy Eka Pratiwi}
\ead{pratiwi.1@iitj.ac.in}

\affiliation[cseiitj]{organization={Department of Computer Science and Engineering, Indian Institute of Technology Jodhpur},
            addressline={NH 62 Nagaur Road, Karwar},
            city={Jodhpur},
            postcode={342037},
            state={Rajasthan},
            country={India}
}

\cortext[cor1]{Corresponding author}

\begin{abstract}
The challenges of graph stream algorithms are twofold. First, each edge needs to be processed only once, and second, it needs to work on highly constrained memory. Diffusion degree is a measure of node centrality that can be calculated (for all nodes) trivially for static graphs using a single Breadth-First Search (BFS). However, keeping track of the Diffusion Degree in a graph stream is nontrivial. The memory requirement for exact calculation is equivalent to keeping the whole graph in memory. The present paper proposes an estimator (or sketch) of diffusion degree for graph streams. We prove the correctness of the proposed sketch and the upper bound of the estimated error. Given $\epsilon, \delta \in (0,1)$, we achieve error below $\epsilon(b_u-a_u)d_u\lambda$ in node $u$ with probability $1-\delta$ by utilizing $O(n\frac1{\epsilon^2}\log{\frac1{\delta}})$ space, where $b_u$ and $a_u$ are the maximum and minimum degrees of neighbors of $u$, $\lambda$ is diffusion probability, and $d_u$ is the degree of node $u$. With the help of this sketch, we propose an algorithm to extract the top-$k$ influencing nodes in the graph stream. Comparative experiments show that the spread of top-$k$ nodes by the proposed graph stream algorithm is equivalent to or better than the spread of top-$k$ nodes extracted by the exact algorithm.
\end{abstract}


\begin{highlights}
\item Diffusion degree is a centrality measure used for influence maximization problem. The present work provide a small space data structure to estimate diffusion degree of every node from a insert only graph stream.
\item We proved the correctness of of the proposed sketch along with the upper bound of error.
\item Given $\epsilon, \delta \in (0,1)$, we achieve error below $\epsilon(b_u-a_u)d_u\lambda$ in node $u$ with probability $1-\delta$ by utilizing $O(n\frac1{\epsilon^2}\log{\frac1{\delta}})$ space, where $b_u$ and $a_u$ are the maximum and minimum degrees of neighbors of $u$, $\lambda$ is diffusion probability, and $d_u$ is the degree of node $u$. 
\item Experimentally we show that the diffusion degree estimate provide a similar or better solution to Influence Maximization problem.
\end{highlights}


\end{frontmatter}

\section{Introduction}
The availability of Big Data allows us to mine large-scale graphs. These graphs are called Complex Networks or Social Networks. 
Traditionally a snapshot of the network is taken as a static graph for developing algorithmic solutions to different problems. 
Most modern social networks are not static; these graphs change with time, new nodes and links are added, and many existing links and nodes are removed over time. Hence, many recent works consider the dynamic nature of the graph to develop solutions that can update efficiently with the change in the graph topology \cite{fppr, ohsaka2016dynamic, Peng2021DynamicIM, li2018}. These graphs are sometimes referred to as temporal graphs. Temporal graphs can be modeled in different ways. One of the techniques is to take multiple snapshots in different timestamps. The temporal network can also be modeled as a dynamic graph where edges appear or disappear from the graph while the whole graph is available for computing problems. However, in many settings, especially the real-world graph generated from online social networks, loading or keeping the whole graph may not be possible due to the velocity of topological changes or the volume of the graph. In this setting, an algorithm must find a solution by observing the graph from the edge stream (high-velocity data) or limited pass of the edge lists (high-volume data). 
For example, in the Twitter user-to-mention network, where a link is formed when one tweet mentions a user, each tweet mentioning someone will add links to the network. That is, the network's topology changes with each tweet. Considering the number of tweets generated in each minute, the change in topology will be faster than the traversal of the graph. Hence, it is best to model it as a graph stream rather than a static or dynamic graph. When graphs are observed as a stream of edges potentially infinite in length, they are called Graph Streams.

Graph streams are mainly used in two cases. Firstly, when there are memory constraints, the graph is processed in a stream of edges instead of loading the entire graph into memory. Secondly, when the graph data comes as a sequence of actions ordered by their timestamps and the velocity of the same is very high. One should note here that the fundamental difference between the dynamic graph model and the streaming graph model is that for the streaming graph, the entire graph is never available to the algorithm, while the dynamic graph has the entire graph available for processing. 
Because of this feature, traversals may be performed in dynamic graphs to find local or global solutions. However, for streaming graph problems, the algorithm must sketch important information to perform a specific task without traversing the graph.

Centrality plays an important role in solving many problems and is extensively used in static settings. For example, edge betweenness is used for community detection; degree, diffusion degree, degree discount, PageRank, and node betweenness are used for finding influential nodes for the problem of influence maximization or viral marketing; PageRank is used for ranking web pages in search engines. Many centrality measures can be efficiently calculated by a single Breadth-First-Search (BFS) (e.g., degree) for all the nodes, while many require exhaustive exploration of the graph (e.g., betweenness). Research has been conducted to update centrality measures for dynamic graph settings (e.g., recalculation of PageRanks after topological changes are done in \cite{10.1007/978-3-030-93409-5_23, Bahmani2010, fppr}), but only a handful number of studies are conducted on estimating different centrality measures for graph streams. In particular, Hayashi et al. \cite{Hayashi2015} have developed a sketching technique to keep track of betweenness for graph streams. To the best of our knowledge, other centrality measures are not explored for graph streams. 

In this current work, we try to answer two questions: (i) `Can we estimate the Diffusion Degree \cite{diffdegree, Pal2014}, a centrality measure, of nodes from graph streams?' and (ii) `Can we use the estimated Diffusion Degree heuristically to solve Influence Maximization problem on streaming graphs, in the same way, it is used for static graphs?' Brief descriptions of the Diffusion Degree and Influence Maximization (IM) problem are provided in Section \ref{sec:preliminaries}.

\begin{table}[!ht]
\centering
\caption{{\bf Important notations used in the paper}}
\label{table:notations}
\begin{tabular}{|p{1cm}|p{6cm}|} 
\hline
Name & Description \\
\hline
$[q]$ & for any integer $q$ it represent $\{1,\dots,q\}$\\ 
$q$ & Number of neighbors to be stored in summary \\
$m$ & Total no. of edges in the graph \\
$d$ & Average degree of graph \\
$d_{in}$ & Average in-degree of graph \\
$d_{u}$ & Degree of node $u$\\
$n$ & Total no. of nodes in the graph \\
$\Gamma(u)$ & Set of neighbors of a node $u$ \\
$\lambda$ & Common diffusion probability \\
$S_q^{(u)}$ & Set of $q$ neighbors of a node $u$\\
$DD_u$ & Diffusion degree of a node $u$ \\
$DDS_u$ & Estimated $DD_u$ in stream\\
\hline
\end{tabular}
\end{table}

\textbf{Our Contribution:} We have developed a streaming algorithm for approximating the diffusion degree measure of nodes from an insert-only (i.e., edges are added to the graph but never deleted from the graph) graph stream. The proposed algorithm is an \textit{online algorithm}, and one can query the data structure at any given time to get the centrality score of any node $u$ in the graph. Instead of storing the full graph, we estimate the diffusion degree of nodes by keeping a few sampled neighbors. Given $\epsilon, \delta \in (0,1)$, we have achieved error below $\epsilon(b_u-a_u)d_u\lambda$ with probability $1-\delta$ by utilizing $O(n\frac1{\epsilon^2}\log{\frac1{\delta}})$ space where $b_u$ and $a_u$ are the maximum and minimum degrees of neighbors of $u$. We also prove the mathematical bound on error. We propose an algorithm to find top-$k$ influencing nodes for the IM problem with the estimated diffusion degree. We show experimentally that the proposed methods work similarly to the static graph algorithms for streaming graphs.

\section{Preliminaries}
\label{sec:preliminaries}

\subsection{Centrality Measure and Diffusion Degree} 
\label{sec:diffdegree} 
Centrality measures are a class of matrices that define the importance of nodes or edges. Node centrality measures are useful to determine a node's importance in a network for a given context. More importantly, centrality measures are useful in ranking the nodes for a given context. It can be defined as: 

\begin{definition}[Centrality]
Centrality Measures are a way to determine which nodes in a graph are the most important. Usually, the centrality measure indicates how central a node is in a network for a particular context \cite{centralitysurvey}.
\end{definition}

There are many centrality measures defined in the literature. Diffusion Degree is one of such centrality measures defined in the context of information diffusion in a network. It is defined as:
 
\begin{definition}[Diffusion Degree]
{The Diffusion Degree is a node's centrality measure developed to find the expected amount of spread by a node in the information diffusion process}.
\end{definition}

It considers not only the degree of the node but also the degree of its neighbors while measuring the centrality score. An Independent Cascade Model (ICM) of diffusion is assumed in the calculations. Although each edge can have different diffusion probabilities, it is assumed to be the same for all the edges in the network. Mathematically, the diffusion degree of node $u$ is defined by
\begin{align}
    \label{diffdeg}
    & DD_u =\lambda \times (d_u + \sum_{i \in \Gamma(u)} d_i) &
\end{align}

where $d_{(.)}$ is the degree of a node $(.)$ and $\Gamma(.)$ returns the set of neighbors of node $(.)$.
\newline

Diffusion degree tends to build a BFS tree for every node $u$ in the graph, containing the nodes that can be influenced by the node $u$. The first layer of this tree consists of neighbors of $u$, the second layer consists of the neighbors of those neighbors, and so on. The diffusion probability usually stays in the order of $10^{-2}$ to $10^{-1}$. Therefore, the chance of activating the nodes in the third layer goes close to zero \cite{chen2009efficient, Pal2014} and can be safely ignored. Thus, the number of nodes in the first two layers is used to decide the centrality value of the node. It was proposed to provide a practically efficient solution for the IM problem on large graphs under the ICM of information diffusion discussed next.

\subsection{Information Diffusion}
Information diffusion is the process by which information spreads through the connections in social graphs. Understanding and modeling the information diffusion process is important in social network analysis. The Independent Cascade Model (ICM), one of the popular information diffusion models, was first proposed by Goldenberg et al. \cite{goldenberg2001talk, goldenberg2001using}. It is a stochastic process that models `word-of-mouth'. An initial seed set is provided as input to the model. Any node can be in two states active or inactive. Active denotes the node is influenced by the information being propagated and inactive indicates the node is not influenced. However, it may happen that an inactive node already received the information from its neighbor but did not choose to change its state. Initially, some of the nodes are chosen as the seed nodes and marked as active. The rest stay inactive. In each time step, each active node tries to influence one of its inactive neighbors. A node converts from inactive to active state with a probability associated with each edge. This parameter of the model is called diffusion probability or propagation probability and is denoted with $\lambda_{(.)}$ in this work. Irrespective of success, the same active node (say $u$) will not try to activate the same neighbor (say $v$) again. However, node $v$ can later be activated by other active nodes in its neighborhood. The influenced nodes (active nodes) further try their neighbors, and so on. This way, the cascade continues until no more nodes can be activated.

\subsection{Influence Maximization}
The problem of Influence Maximization is a well-known problem developed for social networks where the goal is to find $k$ number of seed nodes for which the influence in the network is maximum. This is an NP-hard problem for static \cite{kempe} and dynamic graphs \cite{habiba}. Let a social network be represented with a graph $G = (V, E)$ where vertices or nodes ($V$) are the people or entities and edges ($E$) are their connections or relations. The function $\sigma(S)$ returns the number of nodes influenced by the seed set $S$. The IM problem is to find a set $S$ satisfying the following:
\begin{align}
    & S = \arg_{S\subseteq V} \max_{|S|\leq k} \sigma(S) &
\end{align}
The function $\sigma(.)$ depends on the information diffusion model. This paper uses the ICM, considering that the diffusion degree was originally developed for the same model. In \cite{kempe}, Kempe et al. formulated this problem as a discrete optimization problem and proposed a greedy algorithm that provided $O(1-\frac{1}{e} - \epsilon)$ i.e. about $63\%$ approximation. This is, so far, the best-known approximation of the IM problem in static and dynamic settings.

\subsection{Graph Stream Model}
In this paper, Static graphs are represented with $G(V, E)$ where $V$ is the set of nodes and $E$ is the set of edges. Whereas a graph stream is denoted by $\mathcal{G(V, E)}$ where $\mathcal{V}$ is the set of nodes and $\mathcal{E}$ is the edge stream. 
Each edge $e\in\mathcal{E}$ is associated with a sequence number or timestamp which denotes its position in the stream, and that can be retrieved by function $\tau(.)$. 
An online estimator $<Q_t>(\mathcal{G}, .)$ on $\mathcal{G}$ returns the estimated value of $<Q>$ for a graph $G=(\mathcal{V}, \{e | e\in\mathcal{E} \wedge \tau(e)\leq t\})$.

\section{Estimation of Diffusion Degree on Graph Streams}
\label{sec:main1}
Before discussing the estimation of diffusion degree, let us define a more general problem of estimating centrality measures in a graph stream.

\begin{definition}[Estimation of Centrality Measure in Graph Stream]
The problem of estimating the centrality measure of a given node $u$ from a streaming graph is to find the online estimator $C_t(\mathcal{G}, u)$ over a graph stream $\mathcal{G}$, where $C$ is the desired centrality measure. For example, if one wishes to develop an estimator for degree centrality measure, $C_t(\mathcal{G}, u)$ will return the estimated $d_u$, the degree of the node $u$, of the graph $G=(V=\mathcal{V}, E=\{e | e\in\mathcal{E} \wedge \tau(e)\leq t\})$. In order to get the exact answer, the complete graph out of the stream needs to be kept in the memory. With the estimators, the objective is to reduce memory usage. 
\end{definition}

The particular problem we are addressing is to design an estimator for the diffusion degree centrality.

\subsection{Estimating Diffusion Degree}
\subsubsection{Challenges}
In order to find the diffusion degree $DD_u$ of a node $u$ (\Cref{diffdeg}), we require the degree of node $u$ as well as the sum of the degrees of all of its neighbors. The degree of a node can be sketched by the counter for each node, and it takes only $O(n)$ space for all nodes in the network. The challenge comes when we try to query for the value of the sum of the neighbors. It is required to know who the neighbors of a node are. For the exact solution, the mapping of all the neighbors needs to be kept in the memory, which is equivalent to storing the whole graph if we desire to estimate the diffusion degree of all nodes. A graph stored as an adjacency matrix requires $O(n^2)$ space, while an adjacency list takes $O(n+m)$ storage space. In practical cases, while considering large graphs where $m$ is in the order of $10^6$ or more, the space required (about 229 GB for $10^7$ edges) quickly exceeds typical memory sizes in most RAMs available today. 

\subsubsection{The Solution}
In the proposed approach, we will build a sketch that occupies $O(n\times q)$ memory, where $q$ is the maximum number of neighbors sampled for each nodes. Note that $q$ can be represented with the approximation error and the failure probability; we will discuss these later. The idea is \textit{to store $q$ out of $d_u$ neighbors of a node $u$ instead of all the neighbors.}  Here, $q$ is an input integer value such that $q \ll k_{max}$, the maximum in-degree of any node in the network.

We consider directed graphs while developing the algorithm. This approach can easily be extended for undirected graphs by changing the convention of influence and adding both the nodes of the edge as influencers to each other. Every directed edge $e(v, u)$ has a head node $u$ and a tail node $v$. As a convention, for every edge encountered on the stream, we consider $u$ influences $v$. Hence, for directed graphs, assuming that the node values/ids are integers, we store at most $q$ out of $d_u$ \textit{incoming neighbors} of a node. 
\paragraph{Which $q$ neighbors?}
We select $q$ uniformly randomly sampled with replacement neighbors of $u$. Therefore, each of these $q$ neighbors will have the probability of $\frac1{d_u}$ to be selected to any of the $q$ places. There can be multiple instances of the same node among these $q$ neighbors considering we sample with replacement. Random sampling with the replacement on graph stream is performed with the technique proposed in \cite{rswr}. The procedure is as follows. Initially an array of size $q$, $S_q^{(u)}$ having \textit{null} entries is stored. For every edge in the stream that has $u$ as its head node, we run $q$ independent Bernoulli trials having success probability as $\frac{1}{d_u}$. Among these, $r$ number of trials are successes. We replace $r$ nodes from $S_q^{(u)}$ indexed by their trial numbers with the tail node of the edge. Refer to Figure 2 of \cite{rswr} for the complete algorithm.
We need not know $d_u$ apriori to perform this kind of sampling. Refer to Lemma 2.2 from \cite{rswr} for proof. Note that we cannot use standard reservoir sampling for the proposed approach because this type of sampling is not independent. Without independence, our further proof will not follow. Sampling with replacement guarantees independence between sampled nodes.

\begin{remark}
There are two cases when the number of neighbors in $S_q^{(u)}$ is less than $q$:
\begin{enumerate}
    \item The total number of neighbors of $u$ is less than $q$ in the graph seen so far.
    \item All $q$ places in $S_q^{(u)}$ are not filled by random sampling.
\end{enumerate}
In both cases, consider that the number of neighbors in  $S_q^{(u)}$ is $q'$. We consider $q'$ numbers of neighbors instead of $q$. In the analysis section, we can simply replace $q$ with $q'$ to get the analysis for these cases. For the first case, as we can understand from the formulation, the estimate will be equal to the exact diffusion degree, while it will be approximate in the second case. 
Now that we have our set of neighbors ready, we can use their degrees for the estimation of diffusion degrees.
\end{remark}

\paragraph{The Estimation:}We take the \emph{normalized empirical sum} over the selected neighbors to cover up for the remaining neighbors. 
We thus approximate the diffusion degree value by doing so. Hence, modification to the latter term in \Cref{diffdeg} gives us the estimated value of the diffusion degree of $u$ as:
\begin{align}
    \label{diffdegstream}
   & \widehat{DDS_u} = \lambda \times (d_u + \frac{d_u}{q}\times\sum_{j \in S_q^{(u)}}d_j) &
\end{align}
Note that in the $q$ length array, we do not store the degrees of neighbors of $u$. Instead, we store the node identifier itself. Hence, even for determining a single node's diffusion degree, we need to know the degree of all the $q$ neighbors we are storing. We shall construct an adjacency list-like data structure. For every node $u$, the first element of its list will store the counter of the degree of $u$ ($d_u$). The following elements will store at most $q$ neighbors.

\begin{figure}[!h]
    \includegraphics[width=0.8\linewidth]{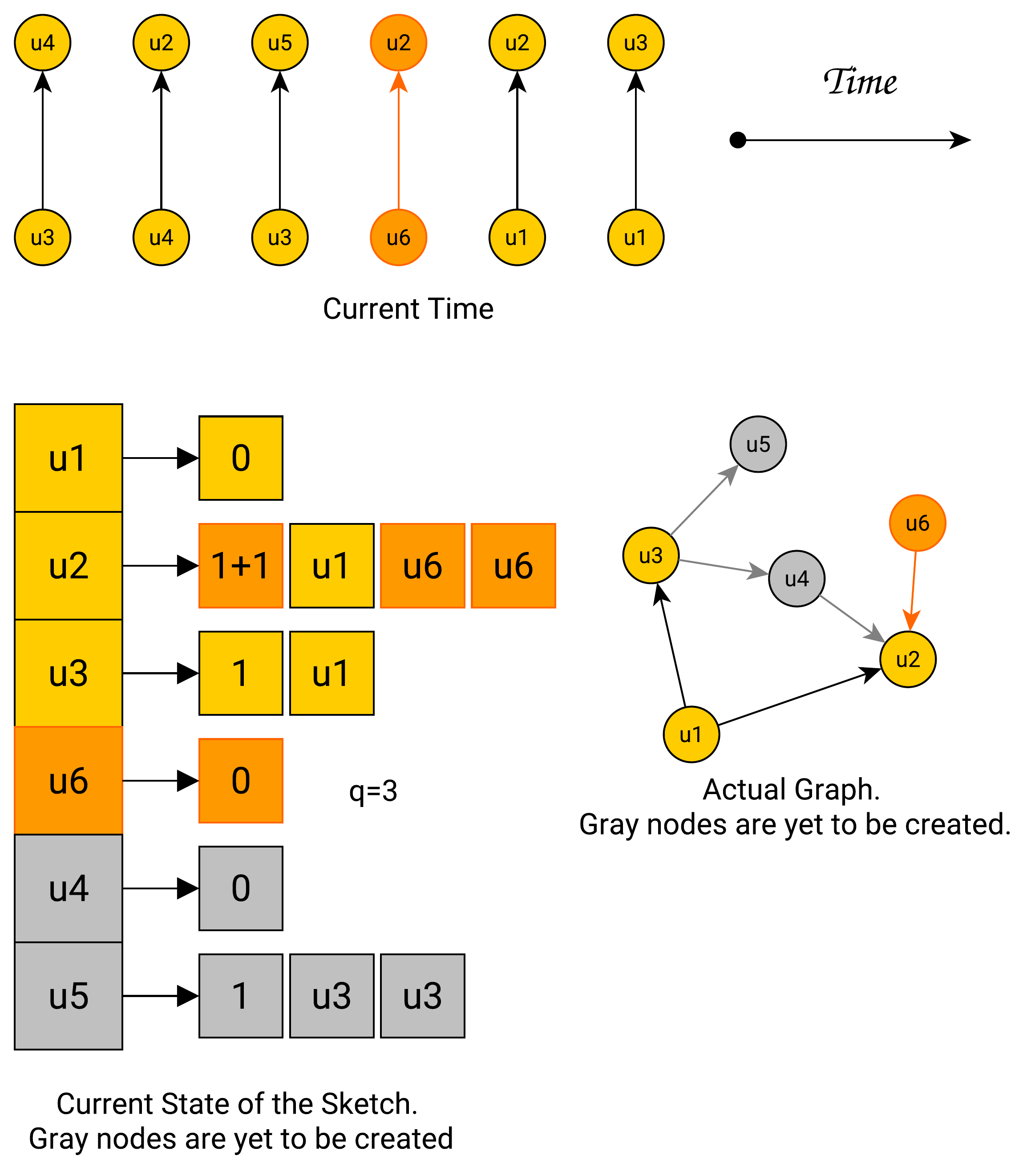}
\caption{Data structure \texttt{ADJ} used to estimate diffusion degrees on edge stream of graph. For every node, we store a (dynamic) list of $q+1$ cells. The first cell stores the degree count for the node, followed by up to $q$ neighbors. $q$ and $\lambda$ are inputs. Every new neighbor is inserted according to random sampling with replacement.}
\label{fig:dds_ds}
\end{figure}

Figure \ref{fig:dds_ds} shows the visualization of the data structure. Initially, all the $q$ neighbors are \textit{null}. Hence, no space is allocated.  The data structure will keep updating with the stream. Specifically, for every new edge ($v, u$),  the $u$ node is queried in the data structure.  Its degree is incremented by 1, and $v$ is added to its list at $r$ positions using uniform random sampling with replacement, as discussed before. Whenever a query for  ${DDS_u}$ is made,  one query is performed to get the list of $u$,  then $O(q+1)$ queries are made to get its degree and traverse through its neighbors, and further $O(q)$ queries are made to get the degrees of its neighbors.  Therefore, a total of $O(2q+1)$ queries are made to get the $DDS$ value of a single node.
\pagebreak
\begin{remark}[Diverse Propagation Probabilities]
    While calculating the diffusion degree, we consider propagation probabilities for all the edges to be equal ($\lambda$), although in real-world cases the propagation probability of each relationship may be different. We consider equal value for $\lambda$ as it allows the development of the theorems discussed next easier without loss of generality. However, note that one can modify the aforementioned data structure to accommodate diverse propagation probabilities. In such a case, instead of keeping the degree of each node, we can keep the expectations at each node. That is, we can keep $\sum_{e(v, u)\in\mathcal{E}}\lambda_{e(v,u)}$ instead of $\sum_{e(v, u)\in\mathcal{E}}1$ and accordingly we can modify the Equation \ref{diffdegstream} as follows.   

\begin{align}
    \label{diffdegstream:lambda}
   \nonumber \widehat{DDS_u} & = (\sum_{e(v, u)\in\mathcal{E}}\lambda_{e(v,u)} +  (\frac{\sum_{e(v, u)\in\mathcal{E}}\lambda_{e(v,u)}}{q} \times\\&\sum_{j \in S_q^{(u)}}\sum_{e(v, j)\in\mathcal{E}}\lambda_{e(v,j)})) &
\end{align} 
\end{remark}

\begin{theorem}\label{thm:correctness}
\textbf{(Correctness)} The expected value of the estimated diffusion degree of any node u equals the diffusion degree value of the same node, i.e., $\mathop{\mathbb{E}}[\widehat{DDS_u}] = DD_u$.

\begin{proof}
\raggedbottom 
From linearity of expectations:
\begin{align*}
    \mathop{\mathbb{E}}[\widehat{DDS_u}] & = \mathop{\mathbb{E}}[\lambda \times ( d_u + \frac{d_u}{q} \times  \sum_{j \in S_q^{(u)}}d_j)] &\\
\end{align*}
Since degrees are independent:
\begin{align}
    \nonumber\mathop{\mathbb{E}}[\widehat{DDS_u}] & = \lambda \times (\mathop{\mathbb{E}}[d_u] +\mathop{\mathbb{E}}[d_u[\frac{1}{q} \times  \sum_{j \in S_q^{(u)}}d_j]) &\\ 
    & = \lambda \times ( d_u + d_u\times\mathop{\mathbb{E}}[(\frac{1}{q} \times \sum_{j \in S_q^{(u)}}d_j)])\label{eq5} 
\end{align}

Here, $\frac{1}{q} \times  \sum_{j \in S_q^{(u)}}d_j$ itself is a random variable. 
Let $X_1$, $X_2$, ..., $X_q$ be independent random variables such that $X_l = d_j$ for $l \in [q]$ and $j \in S_q^{(u)}$ and degree $d_j$ is in the range $[a, b]$. The nodes $j_1$, $j_2$, ..., $j_q$ are sampled from $\Gamma(u)$ uniformly at random. Therefore, $X_l, l \in [q]$ are i.i.d. random variables. 

Therefore, from the expectation of a uniform random variable:\\
\begin{align}
    \mathop{\mathbb{E}}[X_l] & = \mathop{\mathbb{E}}[X] = \frac{1}{d_u}\sum_{i \in \Gamma(u)}d_i ~~~ &\label{iid}\\
    \text{Define:~} \bar{X} & = \frac{1}{q}\sum_{l \in [q]}X_l \label{X_bar}&
\end{align}
From the linearity of expectations, we get:
\begin{align}
    \mathop{\mathbb{E}}[\bar{X}] & = \frac{1}{q}\sum_{l \in [q]}\mathop{\mathbb{E}}[X_l] = \frac{1}{q}\sum_{l \in [q]}(\frac{1}{d_u}\sum_{i \in \Gamma(u)}d_i)& \text{(from \cref{iid})}\nonumber\\
    &= \frac{1}{d_u}\sum_{i \in \Gamma(u)}d_i.\frac{1}{q}\sum_{l \in [q]}(1) & \nonumber
\end{align}
\begin{align}
    &= \frac{1}{d_u}\sum_{i \in \Gamma(u)}d_i.\frac{1}{q}.q &\text{(sum does not have any $l$ term)}\nonumber\\ 
    \mathop{\mathbb{E}}[\bar{X}] &= \frac{1}{d_u}\sum_{i \in \Gamma(u)}d_i \label{exp_X_bar}&
\end{align}
Therefore, after substituting the value of $\bar{X}$ from \Cref{eq5,X_bar}, we get:\\
\begin{align*}
    \mathbb{E}[\widehat{DDS_u}] &= \lambda \times ( d_u + d_u\times\mathop{\mathbb{E}}[\bar{X}]) \\
    &= \lambda \times ( d_u + d_u\times\frac{1}{d_u}\sum_{i \in \Gamma(u)}d_i) && \text{(from \cref{exp_X_bar})}\\
    &= \lambda \times ( d_u + \sum_{i \in \Gamma(u)}d_i)
\end{align*}
Hence, $\mathbb{E}[\widehat{DDS_u}] = DD_u$
\end{proof}
\end{theorem}

\begin{lemma}\label{cor:upperBound}
  For any $\epsilon, \delta \in (0,1)$, with $q=O(\epsilon^{-2}log\frac1{\delta})$, the error between the estimate and the actual value is $|\widehat{DDS_u} - DD_u| \leq \epsilon(b_u-a_u)d_u\lambda$ with probability $1-\delta$.
\end{lemma}
Here, $\epsilon$ is the approximation error and $\delta$ is the error probability. We define $b_u, a_u$ as max degree and min degree among the neighbors of $u$, respectively.
\begin{proof}
Consequently, in Hoeffding's inequality, \\ we can use $\bar{X}$ from \Cref{X_bar} as the estimated random variable. Inserting it in the Hoeffding's inequality, we have:
\begin{align*}
    &\Pr[|\bar{X} - \mathop{\mathbb{E}}[\bar{X}]| \geq \alpha] \leq 2\exp(\frac{-2q\alpha^2}{(b_u-a_u)^2})&
\end{align*}
From \Cref{X_bar,exp_X_bar},
\begin{align*}
    &\Pr[|\frac{1}{q}\sum_{j \in S_q^{(u)}}d_j - \frac{1}{d_u}\sum_{i \in \Gamma(u)}d_i| \geq \alpha] \leq 2\exp(\frac{-2q\alpha^2}{(b_u-a_u)^2})&\\ 
    &\Pr[|\frac{d_u}{q}\sum_{j \in S_q^{(u)}}d_j - \sum_{i \in \Gamma(u)}d_i| \geq \alpha d_u] \leq 2\exp(\frac{-2q\alpha^2}{(b_u-a_u)^2}) &\\
    &\Pr[|\lambda(d_u + \frac{d_u}{q}\sum_{j \in S_q^{(u)}}d_j) - \lambda(d_u + \sum_{i \in \Gamma(u)}d_i)| \geq \alpha d_u\lambda] &\\
    &\hspace{151pt} \leq 2\exp(\frac{-2q\alpha^2}{(b_u-a_u)^2})&
\end{align*}
From \Cref{diffdegstream,diffdeg}
\begin{align*}
    &\Pr[|\widehat{DDS_u} - DD_u| \geq \alpha d_u\lambda] \leq 2\exp(\frac{-2q\alpha^2}{(b_u-a_u)^2})&\\ 
    \end{align*}
For $\epsilon \in (0, 1)$, we put $\alpha = \epsilon(b_u-a_u)$:
\begin{align*}
    &\Pr[|\widehat{DDS_u} - DD_u| \geq \epsilon(b_u-a_u)d_u\lambda] \leq 2\exp(-2q\epsilon^2)&
\end{align*}
Let $\delta \in (0, 1)$ such that $\delta \geq  2\exp(-2q\epsilon^2)$
\begin{align}
    &\Pr[|\widehat{DDS_u} - DD_u| \geq \epsilon(b_u-a_u)d_u\lambda] \leq 2\exp(-2q\epsilon^2) \leq \delta& \label{eq:upperBoundProof}\\
    &\therefore|\widehat{DDS_u} - DD_u| \leq \epsilon(b_u-a_u)d_u\lambda ~\text{ with probability $1-\delta$}& \nonumber
\end{align}
\end{proof}

\begin{remark}[On $b_u$ and $a_u$]
    Note that for scale free network or ultra small world network $b_u$ can be very high. Scale free network follows power law distribution and the maximum degree of a node in the network $k_{max}=k_{min}\times |V|^{\frac1{\gamma}-1}$, where $\gamma$ is the exponent of the power law distribution. However, in the case of scale free network or ultra small world network, the number of high degree hubs are less. We conducted an empirical study with 19 data sets available publicly \cite{nr}. The study reveals that the fraction of nodes having degree more than $\sqrt{|V|}$ is between 0.03\% to 4.7\% except one data set where it is 9.8\%. On the other hand, $a_u$ is equal to or close to 1 for most of the cases. Our empirical study with $19$ networks reveals that the fraction of nodes having degree less than the average degree is at least 61\% and goes up to 84.7\%. Further, we are only concerned about in-degree neighbors, and it is unlikely that a very high degree node is following a relatively less degree node.  
\end{remark}

\subsection{Algorithm} The algorithm is given in \Cref{alg:diffdegstream}. The algorithm maintains an adjacency list \texttt{ADJ} of size $O(n\times q)$ to store the summary of the stream at every time step $t$. Every row $r$ of \texttt{ADJ} stores the summary for the node $u_r$. The first cell of every row counts the degree of $u_r$ over the stream. The remaining $q$ cells store up to $q$ neighbors of $u_r$ with replacement.


\begin{algorithm}
\caption{Diffusion degree estimator: $DDS$(object)}
\footnotesize
\label{alg:diffdegstream}
\begin{algorithmic}[1]
\REQUIRE $q$: max number of neighbors, $\texttt{ADJ}_{t-1}$: adjacency list of size $O(n\times q)$ \\from previous time step, $\texttt{ADJ}_{t'}$: when queried
\STATE \textbf{--- $next(v_t, u_t)$ ---}
\STATE $\texttt{ADJ}_t \gets \texttt{ADJ}_{t-1}$
\STATE{$\texttt{ADJ}_t[u_t][0] \gets \texttt{ADJ}_{t-1}[u_t][0]$ + 1}
\STATE $rswr(v_t, u_t, q)$

\STATE \textbf{--- $query_{t'}(u)$ ---}
\STATE $d_u^{(t')} \gets \texttt{ADJ}_{t'}[u][0]$; $sum \gets 0$; $nCount \gets 0$
\FOR{$i \in [q]$}
    \IF{$\texttt{ADJ}_{t'}[u][i]$ is not $null$}
        \STATE $nCount \gets nCount$ + 1; $sum \gets sum + \texttt{ADJ}_{t'}[\texttt{ADJ}_{t'}[u][i]][0]$
    \ENDIF
\ENDFOR
\IF{$nCount > 0$}
    \RETURN $\lambda$ ($\frac{d_u^{(t')}}{nCount}\times sum +  d_u^{(t')}$)
\ELSE
    \RETURN 0
\ENDIF
\end{algorithmic}
\end{algorithm}

When an edge $e(v_t, u_t)$ from the stream appears at time step $t$, it gets processed in the following way. As discussed in \Cref{sec:main1}, we  consider the nodes' in-degree. Therefore, we consider the head node $u_t$ for processing, and the only row in \texttt{ADJ} updated at time step $t$ is that of node $u_t$. The degree of $u_t$ is incremented by one. After that, the $rswr(.)$ function is called. As discussed before, it performs $q$ Bernoulli trials with probability equivalent to $\frac1{d_{u_t}^{(t)}}$ and stores the result of each one of them in an array of size $q$. Say $r$ of these trials succeed, then we replace the cells at these $r$ indexes with $v_t$. At this point, the ingredients to find the estimate of diffusion degree for $u_t$ are available to us. 

The $query_{t'}(u)$ function finds the estimate of diffusion degree from the stored summary in $\texttt{ADJ}_{t'}[u]$. Notice that $t'$ may be different from $t$, i.e. query can occur anytime, irrespective of the time steps in the stream. To query, we find the count of neighbors with duplicates stored in the row array $\texttt{ADJ}_{t'}[u]$. We add up the degrees of these neighbors and store them in $sum$. When $nCount$ is 0, no neighbors were added, so the estimate becomes 0. Otherwise, we return the normalized empirical sum value as given in Equation \ref{diffdegstream}.

\subsection{Analysis} With this algorithm, our goal is to process the edge stream in limited memory. The above algorithm requires O($n\times q$) memory. If the size of the node data type is $s$, the total memory required is O($nqs$). As for the time complexity, at least O($q$) time is required for every call of $rswr(v_t, u_t, q)$ and that of $query(u)$. In total, the algorithm accesses the data structure \texttt{ADJ} O$(2q+1)$ times.

\section{Influence Maximization Using Diffusion Degree Estimates}

Diffusion degree has been effectively used for Influence Maximization problem. Hence, we would like to address the same problem with the estimated diffusion degree. Let us first define the problem and literature review before providing the solution.

\begin{definition}[Influence Maximization on Graph Streams]
Given a graph stream $\mathcal{G(V, E)}$, the IM problem is to identify the top-$k$ influential nodes seen so far for which the influence in the current topology of the network is maximum. In other words, IM on graph streams is
\begin{align}
   & IM_t(\mathcal{G}, k) = S = \arg_{S\subseteq \mathcal{V}} \max_{|S|\leq k} \sigma(S) &
\end{align}
where $\sigma(S)$ estimates the spread by seed node $S$ on the graph $G=(\mathcal{V}, \{e | e\in\mathcal{E} \wedge \tau(e)\leq t\})$.
\end{definition}

\subsection{Literature Review}
\textit{Influence Maximization} (IM) \cite{kempe2003maximizing, Domingos2001, Richardson2002} under \textit{information diffusion} is a well studied problem in Social Network Analysis. Of the many information diffusion models, the Independent Cascade Model (ICM) and Linear Threshold Model (LTM) \cite{goldenberg2001talk, goldenberg2001using} are the most used. The problem of IM is NP-hard \cite{kempe} under both of these information diffusion models. Researchers tried to use different approximation solutions such as \textit{greedy algorithms} \cite{kempe, Chen} or \textit{heuristic methods} \cite{chen2009efficient, diffdegree, Pal2014}. The majority of the heuristic algorithms use existing (etc., degree, PageRank, etc.) or new (etc., degree discount \cite{chen2009efficient}, diffusion degree \cite{diffdegree, Pal2014} etc.) centrality measures. In general, a higher centrality measure indicates higher influence. One such centrality measure is the Diffusion Degree centrality measure \cite{diffdegree, Pal2014}, designed to work on the ICM of information diffusion. 

As discussed earlier, most modern social networks are not static; these graphs change with time.
IM problem for ever changing graph is to recalculate the top-$k$ influential nodes for each snapshot, assuming they are static. 
The IM problem for dynamic graphs has received significant attention in the research literature recently. Many previous approaches \cite{DBLP:conf/nips/LattanziMNTZ20, DBLP:conf/nips/Monemizadeh20, wang2017real,DBLP:journals/pvldb/OhsakaAYK16, DBLP:conf/sdm/ChenSHX15} in IM for dynamic graphs try to adapt greedy algorithms to the dynamic domain. On the other hand, many use different heuristics to address the same problem \cite{DBLP:conf/icdm/ZhuangSTZS13, DBLP:journals/tkde/YangWPC17, Aggarwal2012OnIN, Liu2017OnTS}. Many works are based upon the usage of Reverse Reachable Set \cite{Borgs2014MaximizingSI, Tang2014InfluenceMN, Tang2015InfluenceMI, Peng2021DynamicIM, ohsaka2016dynamic}. Using heuristics with centrality measures to estimate the ranking of nodes for influence spread has the advantage that every node can be considered independently. Thus, the computational effort required for repeatedly traversing a collection of nodes is saved.

Another stream of research solves the generalized problem of sub-modular optimization over the graph streams. In these works, they consider the influence function $\sigma(S)$ to be sub-modular, i.e., the marginal increase by adding a new element to a smaller set $S$ is at least as high as that of adding a new node to a larger set $S$. According to Theorem 2.2 from \cite{kempe}, the function $\sigma(S)$ used to measure influence spread is a sub-modular function. Some popular works, each building upon the previous algorithms, are Sieve Streaming \cite{sievestreaming}, Sieve Streaming++ \cite{sievestreamingpp}, Three Sieves \cite{threeseives}, etc. These algorithms develop effective summaries of the stream and answer real-time queries using those summaries. However, these generalized methods have to be customized for graph streaming settings. \\

\label{sec:main2} In this section, we design a simple algorithm that uses the diffusion degree estimates to provide an approximate solution for the IM problem. We then analyze the time and space complexities of the complete approach. 

\subsection{The Algorithm} \Cref{alg:imdiffdegstream} ranks nodes using the diffusion degree estimates as the heuristic and outputs a seed set of top-$k$ nodes. It maintains a min-heap \texttt{HEAP} to maintain the top-$k$ influential nodes at every time step $t$. \texttt{HEAP} stores the tuple (estimate, node) for every node added to it. The tuples are arranged in the heap according to their estimate of diffusion degree values. Smaller estimates are evicted over time.

\begin{algorithm}
\caption{IM using Diffusion degree estimates}
\footnotesize
\label{alg:imdiffdegstream}
\begin{algorithmic}[1]
\REQUIRE $K$: Seed set size, $q$: max number of neighbors, $dds$: object of $DDS$, \\ \texttt{HEAP}: min-heap of size $K$
\STATE \textbf{--- $next(e_t)$ ---}
\STATE $(v_t, u_t) \gets e_t; \texttt{HEAP}_t \gets \texttt{HEAP}_{t-1}$
\STATE $dds.next(v_t, u_t)$
\STATE{$estimate \gets dds.query_t(u_t)$}
\IF{$u_t$ is in $\texttt{HEAP}_{t-1}$}
    \STATE update estimate of $u_t$ with $estimate$ in $\texttt{HEAP}_t$
\ELSE 
    \STATE $minEstimate, minNode \gets \texttt{HEAP}_{t-1}.peek()$
    \IF{$estimate > minEstimate$}
        \STATE add $(estimate, u_t)$ to $\texttt{HEAP}_t$; $\texttt{HEAP}_t.pop()$
    \ENDIF
\ENDIF
\STATE \textbf{--- $Query_{t'}()$ ---}
\RETURN $\texttt{HEAP}$

\end{algorithmic}
\end{algorithm}
We first process the edge using $dds.next(v_t, u_t)$. Then, we find the diffusion degree estimate of $u_t$ using $dds.query(u_t)$. While adding to \texttt{HEAP}, we first check if the node $u_t$ was previously added to the heap. If found, we update its estimate with the new $estimate$. Else, if $estimate$ is greater than the smallest estimate, we add $u_t$ to the heap and evict the node with the smallest estimate. 

\subsection{Analysis} The space occupied by the min-heap \texttt{HEAP} is O($k$). Since we are storing the output seed set in the \texttt{HEAP}, we do not consider the memory occupied by it in the space complexity analysis. Therefore, the space complexity of the algorithm is the same as that of $DDS$, i.e., $O(nq)$. The time complexity of the algorithm is $O(m\times k)$. We run over a large edge stream of size $m$, and we process each edge in $O(k)$ time where $k$ time is required to search through the heap ($k\ll m$) (refer \Cref{alg:imdiffdegstream} line 5). However, we search through the heap only if the $estimate > minEstimate$. As the algorithm proceeds over the stream, the $minEstimate$ keeps increasing. Hence, the likelihood of the above comparison being $True$ goes on decreasing, too. So, the amortized analysis of the time complexity would give a value less than O($m\times k$). However, at least O($q$) time is required for every call of the $dds$ method. Therefore, time complexity can also be given as $\Omega(m\times q)$.

\textbf{What is the range of $q$?} Typically, for an approximation algorithm, user inputs are approximation term ($\epsilon$) and error probability ($\delta$). With these two parameters, the user can tune the accuracy of the result while keeping the probability of success ($1-\delta$) above a certain threshold. In this work, $q$ is derived from the error term $\epsilon$ and the probability $\delta$ is as discussed below. 

From \cref{eq:upperBoundProof}, we have $2exp(-2q\epsilon^2) \le \delta$. Therefore, $\epsilon \geq \sqrt{\frac{1}{2q}log\frac{2}{\delta}}$. 
Thus, $\epsilon$ is inversely proportional to $\sqrt{q}$. Therefore, larger values of $q$ will give smaller error values. However, keeping $q$ arbitrarily large will make the \texttt{ADJ} data structure occupy more space. 

\textit{Claim: To gain the space advantage of the algorithm, $q$ should be less than $d_{in}-1$, where $d_{in}$ is the average in-degree of the nodes in the graph stream.}
\begin{proof}
  Considering the size of the datatype of nodes as implicit, the space occupied by a graph stored as an adjacency list is $n+m$. On the other hand, the space occupied by \texttt{ADJ} is given by $n + n(q+1)$.
  In order to get space advantage for the proposed algorithm, the space occupied by \texttt{ADJ} should be less than space required to store the full graph.
  \begin{align*}
      n + n(q+1) <& n+m \\
      \text{Here, } d_{in} = \frac{m}{n}  \\
      n + n(q+1) <& n+nd_{in} \\
      q+2 <& d_{in}+1 \\
      q <& d_{in}-1 
  \end{align*}
\end{proof}
Hence, $q$ could be safely chosen in $[1, d-2]$ according to the desired error and probability.

\begin{remark}
    Note that, for scale free network, even if we choose $q=d_{in}$ the memory usage will be much less than storing the actual graph. The reason is that the high degree nodes and the low degree nodes are not equally distributed around the average degree. That is nodes having degree higher than average degree is far above $d_{in}$ and their count is low. Also, the average degree is comparatively low as there are many nodes with degree very low (1 or 2). Empirical results shows: about $\frac1{3}$ nodes have degree more than the average degree while rest having degree lower than the average degree even though the average degree is at most $\frac{\sqrt{|V|}}{3}$.  
\end{remark}

\section{Comparative Experiment}
Experiments have been performed to check whether the estimated Diffusion Degree and the exact Diffusion Degree perform similarly for IM problems. In other words, the experiments are to validate theoretical claims. Note that the present work does not claim to achieve state-of-the-art results for IM problems. Instead, the IM problem is one of the applications of diffusion degree centrality. As the current work estimates the diffusion degree in a graph stream, it is more relevant to see whether the top-$k$ nodes identified by the estimated diffusion degree produce similar results to that of the top-$k$ nodes identified by the exact diffusion degree.

\begin{figure}[!ht]
    \centering
    \includegraphics[width=\linewidth]{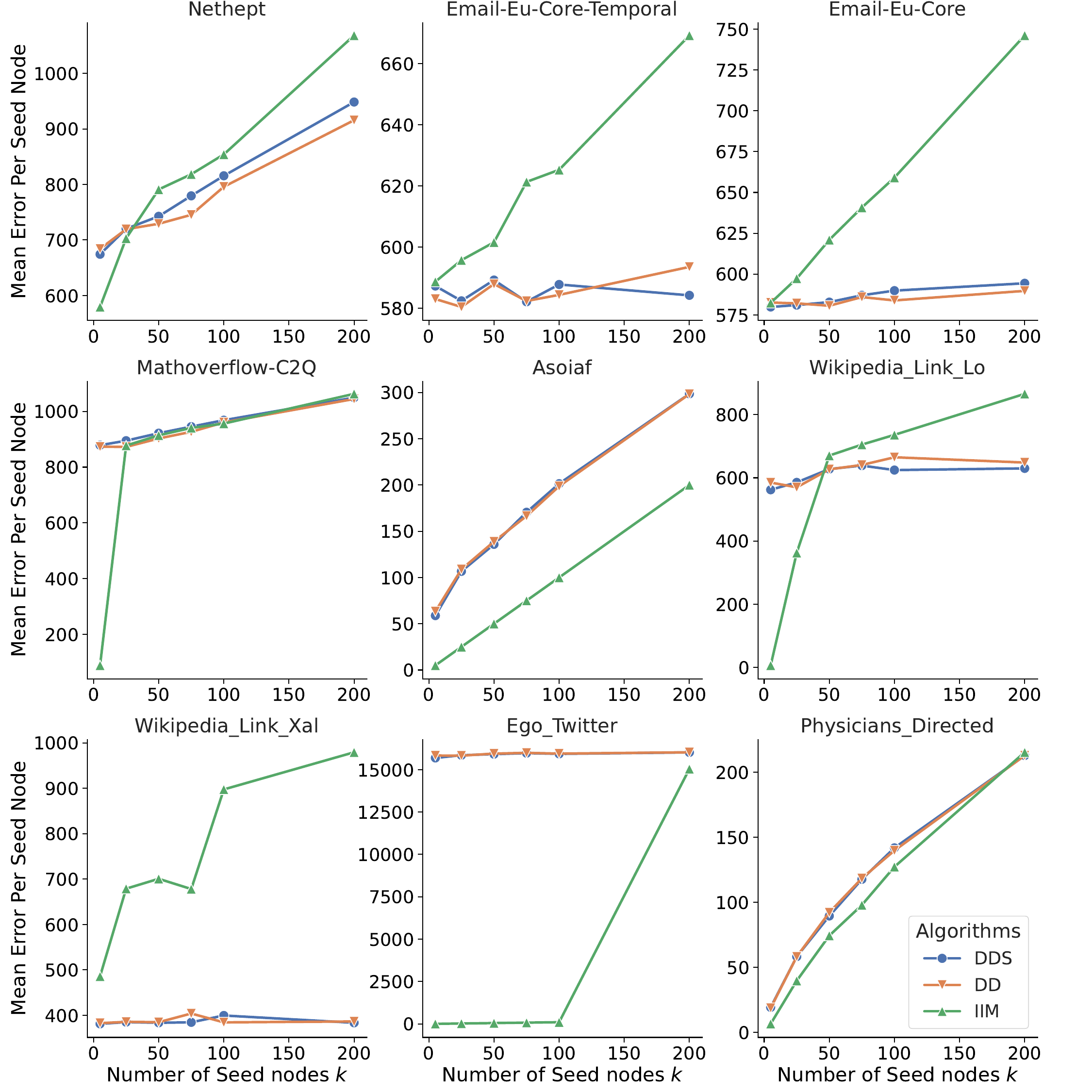}
    \caption{Influence spread with respect to seed set size $K$ on different graphs}
    \label{results}
\end{figure}

\subsection{Experimental Setup}
We used nine datasets in the experiment, as given in \Cref{table:datasets}. All data sets are directed in nature. For diffusion degree, we load the graph as a \textit{networkx} \cite{networkx} object. On the other hand, to simulate the streaming setting, we iterate over a file of directed edges and process each edge at a time. 
The Diffusion Degree (DD) algorithm finds the diffusion degrees of every node and finds the top-$k$ nodes having maximum diffusion degrees. On the other hand for Estimated Diffusion Degree based IM (DDS) we queried for top-$k$ at the end of the stream. We also compared with the Influence Maximization via Martingales (IMM) \cite{IMM} algorithm, and we reused the code available online for the same \footnote{IMM source code: https://sourceforge.net/projects/im-imm/}. To explain the IMM algorithm in a nutshell, it generates a set of nodes called as Reverse-Reachable (RR) set formed by the union of all the nodes that influence a few randomly selected nodes from the graph. The IMM algorithm is optimized in space and time complexity for static network. It is the best performer in the class of sketch-based algorithms \cite{li2018}. Diffusion degree and IMM are deterministic while diffusion degree on streams is stochastic as it depends upon the random sampling. Hence we run the code for 5 times and reported the result of the average.

\begin{table}[h!]
\centering
\caption{Datasets used}
\label{table:datasets}
\begin{tabular}{ |p{3.5cm}|p{1cm}|p{1cm}|p{1cm}|} 
\hline
Name & $n$ & $m$ & $d_{in}$ \\
\hline
Physicians\_directed & 241 & 1098 & 4.556\\
Email-Eu-Core-Temporal & 986 & 24929 & 25.283\\
Email-Eu-Core & 1005 & 25571 & 25.443\\
Mathoverflow-C2Q\cite{mathoverflow} & 16836 & 101329 & 6.018\\
Asoiaf & 796 & 2823 & 3.546 \\
Wikipedia\_Link\_Lo & 3811 & 132837 & 34.856 \\
Wikipedia\_Link\_Xal & 2697 & 232680 & 86.273\\
Ego\_Twitter & 32169 & 643301 & 19.117 \\
Nethept\cite{nethept} & 15229 & 62752 & 4.121\\
\hline
\end{tabular}
\end{table}

For each of the three algorithms and for different values of seed set sizes $k$, we first find the top-$k$ seed set. Then we use NDLib \cite{ndlib1, ndlib2} library for information diffusion to simulate ICM by keeping the seed set as initial active nodes. Every experiment is run for a certain number of simulations, and then the mean over the result of all simulations is reported here as the influence spread for that experiment. The value of $q$ is taken to be $d_{in}-2$ for all the networks. Further, since the diffusion degree stream is stochastic, we run five rounds of the algorithm - every time we find different seed sets and then perform diffusion. Finally, we take the mean over the results of all the runs.

\begin{figure}
    \centering
    \includegraphics[width=\linewidth]{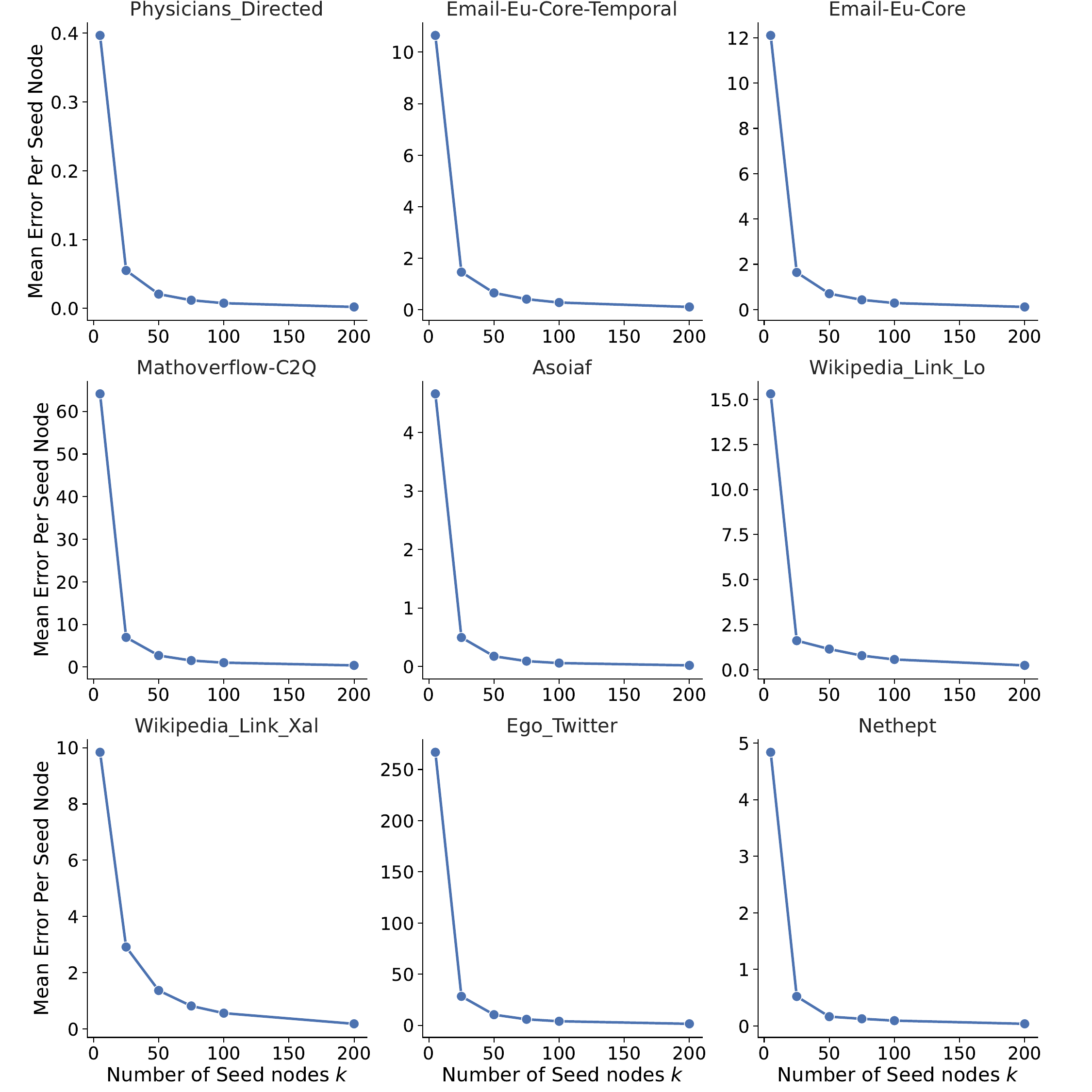}
    \caption{Mean error of DDS for seed nodes}
    \label{fig:mean:error}
\end{figure}

\subsection{Results}
The \Cref{results} shows the experimental results. Each plot shows the estimated spread size of ICM for the seed nodes generated by diffusion degree (DD) diffusion degree on streams (DDS) and Influence Maximization via Martingales (IMM). From the results \Cref{results}, we observe that the influence spread values attained by both DD and DDS algorithms are close to each other for all the data sets used in the experiments. Thereby, it verifies the correctness claim made earlier. 
We also conducted empirical analysis of the error in calculated values of diffusion degree for each nodes. In this experiments we tested for all the seed nodes and mean of the differences in the value of the obtained diffusion degree estimate with the diffusion degree is reported in \Cref{fig:mean:error}. It is observed that as $k$ increases, the mean error gradually decreases for all the data sets.

\begin{figure}[ht!]
    \includegraphics[width=\linewidth]{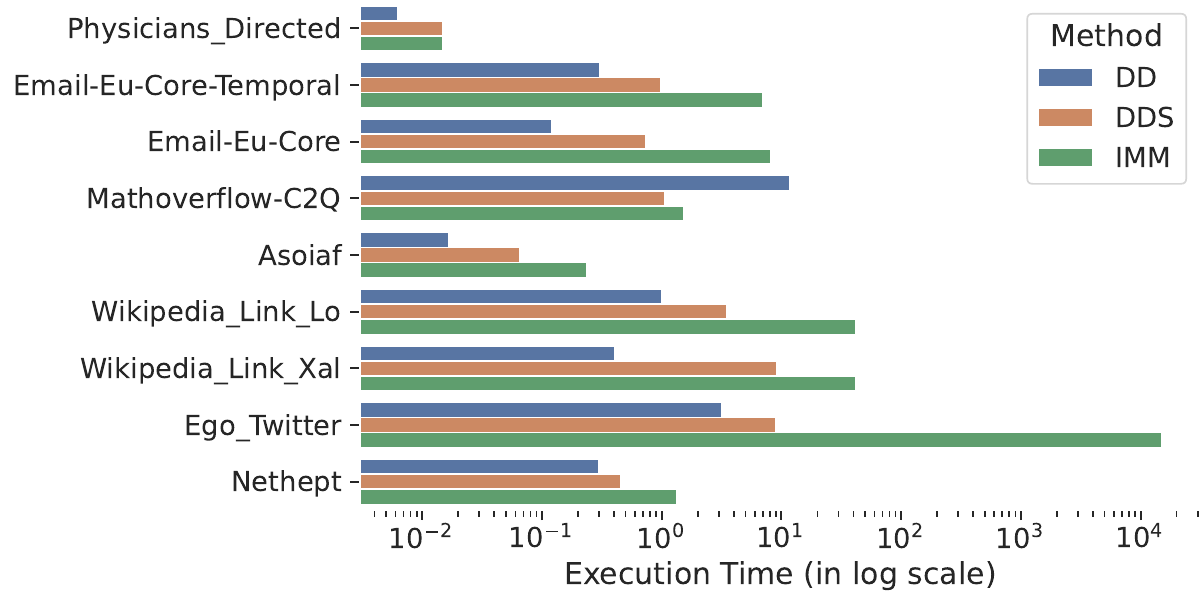}
    \caption{Execution time of different algorithms for different data sets} \label{fig:results2}
\end{figure}

The results in \Cref{fig:results2} show the execution time measured in seconds for different algorithms. The $x$-axis is plotted in logarithmic scale to visually identify the differences as the values for many data sets are too small against the that of the other data sets. In order to keep measurements fair, we have also considered the time required to load the graph into the \textit{networkx} object while measuring time for diffusion degree. The time required to load the graph object would be $O(m)$ and to run diffusion degree over it is $O(n\log n)$. So, the time taken by diffusion degree would be $O(m)$. In the case of diffusion degree on streams, the time required is $O(mk)$. We found that the execution time for DDS is slightly more than that of the DD.

\section{Conclusion}
We proposed an approximation algorithm for calculating diffusion degree from graph stream. Theoretically we shown that the expectation of the estimated values are equal to the exact values of the diffusion degree. The estimated diffusion degree is also used heuristically in finding the top-$k$ influential nodes for Influence Maximization problem. Experimentally, we have shown that the top-$k$ identified by diffusion degree estimate will produce the similar spreading in the network that of the diffusion degree. The proposed algorithm for streaming graph is taking marginally greater time for the whole graph. However, for an infinite stream the algorithm will be able to estimate diffusion degree effectively on multiple queries. Whereas the exact algorithm needs to recalculate for the static graph seen so far. 

Estimating centrality measure in graph stream is least explored in the literature. The present work is in the direction to defining the problem of estimating centrality measure in graph stream.








\section*{Acknowledgments}
Anggy Pratiwi is supported by the Doctoral fellowship in India (DIA) programme of the Ministry of Education, Government of India. 

\section*{Author Contributions}
Suman Kundu contributed in Conceptualization, Methodology, Project administration, Supervision, Visualization and Writing - review \& editing. Vinit Gore involved in Conceptualization, Formal Analysis, Data Curation, Investigation, Methodology, and Writing - original draft. Anggy Pratiwi collaborated in Data curation, Investigation, Methodology, Validation, Resources, and Writing - review \& editing.

\bibliography{main}

\begin{thebibliography}{10}
\expandafter\ifx\csname url\endcsname\relax
  \def\url#1{\texttt{#1}}\fi
\expandafter\ifx\csname urlprefix\endcsname\relax\def\urlprefix{URL }\fi
\expandafter\ifx\csname href\endcsname\relax
  \def\href#1#2{#2} \def\path#1{#1}\fi

\bibitem{fppr}
K.~Suman, R.~Pashikanti~P, Fppr: Fast pessimistic (dynamic) pagerank to update
  pagerank in evolving directed graphs on network changes, in: Social Network
  Analysis and Mining Journal, 2022.

\bibitem{ohsaka2016dynamic}
N.~Ohsaka, T.~Akiba, Y.~Yoshida, K.-i. Kawarabayashi, Dynamic influence
  analysis in evolving networks, Proceedings of the VLDB Endowment 9~(12)
  (2016) 1077--1088.

\bibitem{Peng2021DynamicIM}
B.~Peng, Dynamic influence maximization, in: NeurIPS, 2021.

\bibitem{li2018}
Y.~Li, J.~Fan, Y.~Wang, K.-L. Tan, Influence maximization on social graphs: A
  survey, IEEE Transactions on Knowledge and Data Engineering 30~(10) (2018)
  1852--1872.

\bibitem{10.1007/978-3-030-93409-5_23}
R.~Parjanya, S.~Kundu, {FPPR: Fast Pessimistic PageRank for Dynamic Directed
  Graphs}, in: R.~M. Benito, C.~Cherifi, H.~Cherifi, E.~Moro, L.~M. Rocha,
  M.~Sales-Pardo (Eds.), Complex Networks {\&} Their Applications X, Springer
  International Publishing, Cham, 2022, pp. 271--281.

\bibitem{Bahmani2010}
B.~Bahmani, A.~Chowdhury, A.~Goel, {Fast incremental and personalized
  PageRank}, Proceedings of the VLDB Endowment 4~(3) (2010) 173--184.
\newblock \href {http://arxiv.org/abs/1006.2880} {\path{arXiv:1006.2880}},
  \href {https://doi.org/10.14778/1929861.1929864}
  {\path{doi:10.14778/1929861.1929864}}.

\bibitem{Hayashi2015}
T.~Hayashi, T.~Akiba, Y.~Yoshida, {Fully dynamic betweenness centrality
  maintenance on massive networks}, Proceedings of the VLDB Endowment 9~(2)
  (2015) 48--59.
\newblock \href {https://doi.org/10.14778/2850578.2850580}
  {\path{doi:10.14778/2850578.2850580}}.

\bibitem{diffdegree}
S.~Kundu, C.~Murthy, S.~K. Pal, A new centrality measure for influence
  maximization in social networks, in: International conference on pattern
  recognition and machine intelligence, Springer, 2011, pp. 242--247.

\bibitem{Pal2014}
S.~K. Pal, S.~Kundu, C.~A. Murthy, Centrality measures, upper bound, and
  influence maximization in large scale directed social networks, Fundamenta
  Informaticae 130 (2014) 317--342.
\newblock \href {https://doi.org/10.3233/FI-2014-994}
  {\path{doi:10.3233/FI-2014-994}}.

\bibitem{centralitysurvey}
A.~Saxena, S.~Iyengar, Centrality measures in complex networks: A survey, arXiv
  preprint arXiv:2011.07190 (2020).

\bibitem{chen2009efficient}
W.~Chen, Y.~Wang, S.~Yang, Efficient influence maximization in social networks,
  in: Proceedings of the 15th ACM SIGKDD International Conference on Knowledge
  Discovery and Data Mining, 2009, pp. 199--208.

\bibitem{goldenberg2001talk}
J.~Goldenberg, B.~Libai, E.~Muller, Talk of the network: A complex systems look
  at the underlying process of word-of-mouth, Marketing letters 12~(3) (2001)
  211--223.

\bibitem{goldenberg2001using}
J.~Goldenberg, B.~Libai, E.~Muller, Using complex systems analysis to advance
  marketing theory development: Modeling heterogeneity effects on new product
  growth through stochastic cellular automata, Academy of Marketing Science
  Review 9~(3) (2001) 1--18.

\bibitem{kempe}
D.~Kempe, J.~Kleinberg, {\'E}.~Tardos, Maximizing the spread of influence
  through a social network, in: Proceedings of the ninth ACM SIGKDD
  International Conference on Knowledge Discovery and Data Mining, 2003, pp.
  137--146.

\bibitem{habiba}
T.~B.~W. Habiba, T.~Y. Berger-Wolf,
  \href{http://archive.dimacs.rutgers.edu/TechnicalReports/abstracts/2007/2007-20.html}{Maximizing
  the extent of spread in a dynamic network}, Tech. rep., DIMACS (2007).
\newline\urlprefix\url{http://archive.dimacs.rutgers.edu/TechnicalReports/abstracts/2007/2007-20.html}

\bibitem{rswr}
B.-H. Park, G.~Ostrouchov, N.~F. Samatova, Sampling streaming data with
  replacement, Computational statistics \& data analysis 52~(2) (2007)
  750--762.

\bibitem{nr}
R.~A. Rossi, N.~K. Ahmed, \href{https://networkrepository.com}{The network data
  repository with interactive graph analytics and visualization}, in: AAAI,
  2015.
\newline\urlprefix\url{https://networkrepository.com}

\bibitem{kempe2003maximizing}
D.~Kempe, J.~Kleinberg, {\'E}.~Tardos, Maximizing the spread of influence
  through a social network, in: Proceedings of the ninth ACM SIGKDD
  International Conference on Knowledge Discovery and Data Mining, 2003, pp.
  137--146.

\bibitem{Domingos2001}
P.~Domingos, M.~Richardson, Mining the network value of customers, in:
  Processings of the 7th ACM SIGKDD International Conference on Knowledge
  Discovery and Data Mining, ACM, 2001, pp. 57--66.

\bibitem{Richardson2002}
M.~Richardson, P.~Domingos, Mining knowledge-sharing sites for viral marketing,
  in: Proc. of the 8th ACM SIGKDD International Conference on Knowledge
  Discovery and Data Mining, ACM Press, 2002, pp. 61--70.

\bibitem{Chen}
W.~Chen, C.~Wang, Y.~Wang, Scalable influence maximization for prevalent viral
  marketing in large-scale social networks, in: Proceedings of the 16th ACM
  SIGKDD Conference on Knowledge Discovery and Data Mining (KDD'2010),
  Washington DC, U.S.A., 2010.

\bibitem{DBLP:conf/nips/LattanziMNTZ20}
S.~Lattanzi, S.~Mitrovic, A.~Norouzi{-}Fard, J.~Tarnawski, M.~Zadimoghaddam,
  Fully dynamic algorithm for constrained submodular optimization, in:
  H.~Larochelle, M.~Ranzato, R.~Hadsell, M.~Balcan, H.~Lin (Eds.), Advances in
  Neural Information Processing Systems 33: Annual Conference on Neural
  Information Processing Systems, (NeurIPS 2020), 2020.

\bibitem{DBLP:conf/nips/Monemizadeh20}
M.~Monemizadeh, Dynamic submodular maximization, in: H.~Larochelle, M.~Ranzato,
  R.~Hadsell, M.~Balcan, H.~Lin (Eds.), Advances in Neural Information
  Processing Systems 33: Annual Conference on Neural Information Processing
  Systems, (NeurIPS 2020), 2020.

\bibitem{wang2017real}
Y.~Wang, Q.~Fan, Y.~Li, K.-L. Tan, Real-time influence maximization on dynamic
  social streams, arXiv preprint arXiv:1702.01586 (2017).

\bibitem{DBLP:journals/pvldb/OhsakaAYK16}
N.~Ohsaka, T.~Akiba, Y.~Yoshida, K.~Kawarabayashi, Dynamic influence analysis
  in evolving networks, Proc. {VLDB} Endow. 9~(12) (2016) 1077--1088.
\newblock \href {https://doi.org/10.14778/2994509.2994525}
  {\path{doi:10.14778/2994509.2994525}}.

\bibitem{DBLP:conf/sdm/ChenSHX15}
X.~Chen, G.~Song, X.~He, K.~Xie, On influential nodes tracking in dynamic
  social networks, in: S.~Venkatasubramanian, J.~Ye (Eds.), Proceedings of the
  2015 {SIAM} International Conference on Data Mining, Vancouver, BC, Canada,
  April 30 - May 2, 2015, {SIAM}, 2015, pp. 613--621.
\newblock \href {https://doi.org/10.1137/1.9781611974010.69}
  {\path{doi:10.1137/1.9781611974010.69}}.

\bibitem{DBLP:conf/icdm/ZhuangSTZS13}
H.~Zhuang, Y.~Sun, J.~Tang, J.~Zhang, X.~Sun, Influence maximization in dynamic
  social networks, in: H.~Xiong, G.~Karypis, B.~M. Thuraisingham, D.~J. Cook,
  X.~Wu (Eds.), 2013 {IEEE} 13th International Conference on Data Mining,
  Dallas, TX, USA, December 7-10, 2013, {IEEE} Computer Society, 2013, pp.
  1313--1318.
\newblock \href {https://doi.org/10.1109/ICDM.2013.145}
  {\path{doi:10.1109/ICDM.2013.145}}.

\bibitem{DBLP:journals/tkde/YangWPC17}
Y.~Yang, Z.~Wang, J.~Pei, E.~Chen, Tracking influential individuals in dynamic
  networks, {IEEE} Transactions on Knowledge Data Engineering 29~(11) (2017)
  2615--2628.
\newblock \href {https://doi.org/10.1109/TKDE.2017.2734667}
  {\path{doi:10.1109/TKDE.2017.2734667}}.

\bibitem{Aggarwal2012OnIN}
C.~C. Aggarwal, S.~Lin, P.~S. Yu, On influential node discovery in dynamic
  social networks, in: Proceedings of the 12th {SIAM} International Conference
  on Data Mining, {SIAM} / Omnipress, 2012, pp. 636--647.
\newblock \href {https://doi.org/10.1137/1.9781611972825.55}
  {\path{doi:10.1137/1.9781611972825.55}}.

\bibitem{Liu2017OnTS}
X.~Liu, X.~Liao, S.~Li, J.~Zhang, L.~Shao, C.~Huang, L.~Xiao, On the shoulders
  of giants: Incremental influence maximization in evolving social networks,
  ArXiv abs/1508.00987 (2017).

\bibitem{Borgs2014MaximizingSI}
C.~Borgs, M.~Brautbar, J.~T. Chayes, B.~Lucier, Maximizing social influence in
  nearly optimal time, in: SODA, 2014.

\bibitem{Tang2014InfluenceMN}
Y.~Tang, X.~Xiao, Y.~Shi, Influence maximization: near-optimal time complexity
  meets practical efficiency, Proceedings of the 2014 ACM SIGMOD International
  Conference on Management of Data (2014).

\bibitem{Tang2015InfluenceMI}
Y.~Tang, Y.~Shi, X.~Xiao, Influence maximization in near-linear time: A
  martingale approach, Proceedings of the 2015 ACM SIGMOD International
  Conference on Management of Data (2015).

\bibitem{sievestreaming}
A.~Badanidiyuru, B.~Mirzasoleiman, A.~Karbasi, A.~Krause, Streaming submodular
  maximization: Massive data summarization on the fly, in: Proceedings of the
  20th ACM SIGKDD International Conference on Knowledge Discovery and Data
  Mining, 2014, pp. 671--680.

\bibitem{sievestreamingpp}
E.~Kazemi, M.~Mitrovic, M.~Zadimoghaddam, S.~Lattanzi, A.~Karbasi, Submodular
  streaming in all its glory: Tight approximation, minimum memory and low
  adaptive complexity, in: International Conference on Machine Learning, PMLR,
  2019, pp. 3311--3320.

\bibitem{threeseives}
S.~Buschj{\"a}ger, P.-J. Honysz, L.~Pfahler, K.~Morik, Very fast streaming
  submodular function maximization, in: Joint European Conference on Machine
  Learning and Knowledge Discovery in Databases, Springer, 2021, pp. 151--166.

\bibitem{networkx}
A.~Hagberg, P.~Swart, D.~S~Chult, Exploring network structure, dynamics, and
  function using networkx, Tech. rep., Los Alamos National Lab.(LANL), Los
  Alamos, NM (United States) (2008).

\bibitem{IMM}
Y.~Tang, Y.~Shi, X.~Xiao, Influencemaximizationinnear-lineartime:amartingale
  approach, in: Proceedings of the 2015 ACM SIGMOD International Conference on
  Management of Data, 2015, p. 1539–1554.

\bibitem{mathoverflow}
A.~Paranjape, A.~R. Benson, J.~Leskovec, Motifs in temporal networks, in:
  Proceedings of the tenth ACM International Conference on Web Search and Data
  Mining, 2017, pp. 601--610.

\bibitem{nethept}
N.~Chen, On the approximability of influence in social networks, SIAM Journal
  on Discrete Mathematics 23~(3) (2009) 1400--1415.

\bibitem{ndlib1}
G.~Rossetti, L.~Milli, S.~Rinzivillo, A.~S{\^\i}rbu, D.~Pedreschi,
  F.~Giannotti, Ndlib: a python library to model and analyze diffusion
  processes over complex networks, International Journal of Data Science and
  Analytics 5~(1) (2018) 61--79.

\bibitem{ndlib2}
G.~Rossetti, L.~Milli, S.~Rinzivillo, A.~Sirbu, D.~Pedreschi, F.~Giannotti,
  Ndlib: Studying network diffusion dynamics, in: 2017 IEEE International
  Conference on Data Science and Advanced Analytics (DSAA), IEEE, 2017, pp.
  155--164.

\end{thebibliography}

\end{document}